\documentclass[twocolumn]{aastex63}

\usepackage{appendix}

\received{February 4, 2021}
\revised{\today}

\submitjournal{Astrophysical Journal}

\shorttitle{Radioactive decay of $r$-process nuclei}
\shortauthors{Sprouse et al.}

\graphicspath{{./}{figs/}}

\begin{document}

\hspace{5.2in} \mbox{LA-UR-21-20819}

\title{Isochronic evolution and the radioactive decay of $r$-process nuclei}

\correspondingauthor{T.~M. Sprouse}
\email{tmsprouse@lanl.gov}

\author[0000-0002-4375-4369]{T.~M. Sprouse}
\affiliation{Theoretical Division, Los Alamos National Laboratory, Los Alamos, NM, 87545, USA}
\affiliation{Center for Theoretical Astrophysics, Los Alamos National Laboratory, Los Alamos, NM, 87545, USA}

\author[0000-0002-0637-0753]{G.~Wendell Misch}
\affiliation{Theoretical Division, Los Alamos National Laboratory, Los Alamos, NM, 87545, USA}
\affiliation{Center for Theoretical Astrophysics, Los Alamos National Laboratory, Los Alamos, NM, 87545, USA}
\affiliation{Joint Institute for Nuclear Astrophysics - Center for the Evolution of the Elements, USA}

\author[0000-0002-9950-9688]{M.~R. Mumpower}
\affiliation{Theoretical Division, Los Alamos National Laboratory, Los Alamos, NM, 87545, USA}
\affiliation{Center for Theoretical Astrophysics, Los Alamos National Laboratory, Los Alamos, NM, 87545, USA}
\affiliation{Joint Institute for Nuclear Astrophysics - Center for the Evolution of the Elements, USA}

\begin{abstract}

We report on the creation and application of a novel decay network that uses the latest data from experiment and evaluation. 
We use the network to simulate the late-time phase of the rapid neutron capture ($r$) process. 
In this epoch, the bulk of nuclear reactions, such as radiative capture, have ceased and nuclear decays are the dominant transmutation channels. 
We find that the decay from short-lived to long-lived species naturally leads to an isochronic evolution in which nuclei with similar half-lives are populated at the same time.  
We consider random perturbations along each isobaric chain to initial solar-like $r$-process compositions to demonstrate the isochronic nature of the late-time phase of the $r$-process. 
Our analysis shows that detailed knowledge of the final isotopic composition allows for the prediction of late-time evolution with a high degree of confidence despite uncertainties that exist in astrophysical conditions and the nuclear physics properties of the most neutron-rich nuclei. 
We provide the time-dependent nuclear composition in the Appendix as supplemental material. 

\end{abstract}


\section{Introduction} \label{sec:intro}
The physical mechanisms for the creation of the heavy elements range from the slow ($s$) to the rapid ($r$) capture of free neutrons on nuclei among the stars \citep{Cameron1957, Burbidge1957}. 
The $s$ process synthesizes heavier nuclides in environments where the timescale for neutron capture is much slower than the $\beta$-decay half-lives of participating nuclei \citep{Seeger1965, Sneden2008}. 
In contrast, the $r$ process initially builds heavier nuclei on a fast timescale relative to the short $\beta$-decay half-lives of neutron-rich species \citep{Mathews1990, Freiburghaus1999}. 
As the available neutrons in the environment are depleted, nuclei decay back to stability, releasing energy that may generate an observable signal \citep{Li1998, Metzger2010, Abbott2017a}. 
The $r$ process is the only nucleosynthesis process capable of producing the heaviest elements found in nature \citep{Goriely2015, Zhu2018, Holmbeck2019, Giuliani2020, Wang2020}. 

Despite its importance, a complete description of the $r$ process remains the subject of continued research, as reviewed in \cite{Horowitz2019, Kajino2019,Arnould2020}. 
One of the greatest barriers to constructing a more complete picture of the $r$ process is the large uncertainties in properties of the thousands of participating nuclei \citep{Arnould2007}. 
Reducing uncertainties is complicated by the fact that the missing data required to model the $r$ process primarily reside in unexplored regions of the nuclear chart where nuclei exist for only fractions of a second before decaying \citep{Hosmer2005, Sun2008,McDonnell2015,Schunck2015}.

Sensitivity studies isolate the nuclei whose uncertainties may have the highest impact on $r$-process simulations and provide researchers with a targeted means to focus future measurement campaigns \citep{Aprahamian2014,Surman2014,Mumpower2014, Mumpower2016r}. 
Studies of decay properties (half-lives and branching ratios) have identified select areas near closed shells and far from stable isotopes \citep{Surman2015}. 
Experimental efforts to study these properties have improved predictions from $r$-process simulations \citep{Spyrou2016, Dillmann2018, Lyons2019, Wu2020} but have not yet reached the most influential nuclei \citep{Mumpower2014}. 
Other nuclear properties relevant to the $r$ process---e.g.~binding energies \citep{Orford2018, Tang2020, Vilen2020} and cross sections \citep{Liddick2016, Spyrou2017, Bliss2017}---present their own obstacles to measurement \citep{Cowan2019}. 

Where experimental or evaluated data \citep{Wang2017, NuBase2016, ENDFB8} are not available, nuclear theory must fill in the gaps all the way out to the neutron dripline \citep{Erler2012, Neufcourt2020, Tsunoda2020}. 
Shell model calculations supply accurate estimates for nuclei near closed shells \citep{Pinedo1999, Cuenca2007,Suzuki2012, Zhi2013}. 
Outside these regions, global calculations such as the Finite-Range Droplet Model can supply needed nuclear quantities \citep{Moller2016, Mumpower2016, Mumpower2018, Moller2019, Mumpower2020,Vassh2020}. 
Modern microscopic calculations have also made significant advancements in the description of properties of heavy nuclei \citep{Marketin2016, Shafer2016, Baldo2017, Bulgac+18, Ney2020}. Despite these broad and varied theoretical techniques, much remains to be explored regarding the structure of neutron-rich nuclei \citep{Giuliani2018,Vassh2019, Sprouse2020a}. 

In this work, we focus on understanding the late-time radioactive decay of $r$-process nuclei that occurs after the capture of free neutrons has completed. 
This epoch of the $r$ process is notable from a modeling perspective as it involves nuclei with the most experimentally verified properties among $r$-process participants \citep{RA2020, Korobkin2020}.
We simulate the post-neutron-capture phase using a novel decay network, Jade, that handles nuclear decays and transitions between excited states as detailed in Sec.~\ref{sec:meth}. 
We discuss the isochronic nature of the evolution (Sec.~\ref{sec:results}) that arises when neutron-rich nuclei with short half-lives decay into longer-lived products. 
We provide snapshots of the nuclear composition as a function of time for use in future studies. 

\section{Methods} \label{sec:meth}
\subsection{Modeling Radioactive Decay Nucleosynthesis}
In a system composed of atomic nuclei, and for which the system's temperature, density, and related quantities are well defined, the nuclear abundances follow a generalized set of equations as variously presented, e.g.,  in  \cite{Hix1999,Lippuner2017,Sprouse2020b}. When the astrophysical conditions are not sufficient for nuclear reactions to proceed, the network equations simplify to 
\begin{equation}\label{eq:decay_net}
\frac{dY_{i,j}}{dt} = -\lambda_{i,j} (t) Y_{i,j}(t) +\sum\limits_{(k,l)\neq (i,j)} P_{k,l}^{i,j}\lambda_{k,l} (t) Y_{k,l}  \ ,  
\end{equation}
where 
$i$ indexes unique nuclear species,
$j$ indexes the ground and long-lived excited states of species $i$,
$\lambda_{i,j}(t)$ is is the total decay rate of species $Y_{i,j}$ and allowed to depend generically on time, $t$, and 
the summation over ($k,l$) is taken over all species which decay into species ($i,j$) with branching ratio $P_{k,l}^{i,j}$. 

A system described by Eq.~\ref{eq:decay_net} is a linear, ordinary, first-order, and homogeneous system of differential equations. It may be solved as an Initial Value Problem using any number of numerical and/or exact techniques. Under specific conditions, the Bateman Equations \citep{bateman1910} and its generalizations \citep{furuta1987,wilson1998} provide an exact, analytical solution for the $Y_{i,j}(t)$, but these approaches are not always appropriate, as cancellation errors can arise whenever $\lambda_{i,j} \approx \lambda_{k,l},$ and they are not applicable for systems in which both the forward $(i,j) \rightarrow (k,l)$ and reverse $(k,l) \rightarrow (i,j)$ transitions between nuclear states $(i,j)$ and $(k,l)$ are allowed to occur \citep{thomas1994}, although alternate constructions have been proposed that aim to overcome these limitations \citep{cetnar2006}.

Alternatively, the matrix exponential offers a natural method for solving Eq.~\ref{eq:decay_net}. If we construct the rate matrix, $\Lambda$, with entries
\begin{equation}
    {\Lambda_{m,n} = -\lambda_m \delta_{m,n} + \sum_{n} P_{n}^{m}\lambda_n (1-\delta_{m,n})} \ ,
\end{equation} where $m$ and $n$ enumerate the unique ($i,j$) pairs in the decay network, and the sum over $n$ is taken over all species which decay to produce species $m$. The $\delta_{m,n}$ is the Kronecker $\delta$ function and separates the negative diagonal elements of $\Lambda$, representing the \textit{destruction} of nuclei, from the positive off-diagonal elements of $\Lambda$, representing the \textit{production} of nuclei. With the matrix $\Lambda$ thus defined, Eq.~\ref{eq:decay_net} may be recast as
\begin{equation}\label{eq:decay_net_mat}
    \frac{d\vec{Y}}{dt} = \Lambda \vec{Y } \ ,
\end{equation}
where $\vec{Y}$ is a column vector of nuclear abundances whose $m\textrm{-th}$ entry is the abundance of nuclear species $m.$

If we allow for some time-dependence to the matrix $\Lambda$, for example temperature-dependent transition rates in a non-constant thermal environment \citep{Ward1980,Coc2000,Misch2020, Misch2021}, then the solution can be explicitly obtained using the matrix exponential as
\begin{equation}\label{eq:full_decay_soln}
    \vec Y(t) =  \exp{\left(\int_{t_0}^{t}\Lambda (t') dt'\right)} \vec Y(t_0) \ ,
\end{equation}
where the integral is performed element-wise on $\Lambda$, and the matrix exponential is defined for any square matrix $A$ as the infinite series 
\begin{equation}\label{eq:mat_exp}
    \exp{A} = \sum\limits_{i=0}^{\infty} \frac{A^i}{i!} \ .
\end{equation}
If we restrict ourselves to problems where $\Lambda$ is constant in time, then the solution to Eq.~\ref{eq:decay_net_mat} reduces to
\begin{equation}\label{eq:decay_soln}
    \vec Y(t) =  \exp{\Big((t-t_0)\cdot\Lambda \Big)} \vec Y(t_0) \ .
\end{equation} 

In exchange for greater flexibility (compared to methods aimed at exact solutions), approaches based on Eq.~\ref{eq:decay_soln} suffer from well-known difficulties associated with the evaluation of the matrix exponential \citep{thomas1994,yamamoto2007}. For example, if Eq.~\ref{eq:mat_exp} is invoked directly, and the summation truncated to a finite number of terms, convergence can be poor, and the accuracy of the solution is subject to roundoff errors associated with summing a large number of negative and positive terms. In general, the restriction to smaller timesteps may be associated with greater numerical precision overall, in exchange for a larger number of timesteps being required to evolve a system over a given period of time.  As we detail in Sec.~\ref{sec:jade}, we adopt an adaptive refinement procedure based on this general strategy.

\subsection{The Jade Decay Network Solver}\label{sec:jade}
In this section, we provide an overview of the Jade decay network solver. We discuss here only the means by which the abundances $Y_{i,j}$ associated with a nuclear composition may be evolved from an initial state  at time $t=t_0$ to a final time $t=t_f,$ as other quantities of interest to the $r$ process evolution, such as total and effective nuclear self-heating rates, can be directly obtained once the abundances are known. 

Jade begins by enumerating a discrete list of timesteps $t_0,t_1,\dots,t_f$ for which we wish to solve for the abundances $\vec Y.$ At each timestep  $t_i,$ the abundances can be evolved from $t_i$ to $t_{i+1}$ by evaluating Eq.~\ref{eq:decay_soln}, namely
\begin{equation}
    \vec{Y}(t_{i+1}) = \exp\Big ( (t_{i+1}-t_i)\cdot\Lambda \Big ) \vec{Y} (t_i)
\end{equation}
For the matrix exponential, we incorporate the current SciPy implementation \citep{Virtanen2020} which is based on a scaling-and-squaring technique \citep{AlMohy2010} that has been successfully applied in a variety of decay network studies similar to the type considered here \citep{moler2003,pusa2010,gauld2011}. In the interest of preserving numerical accuracy, it is necessary to limit the total timestep size, $\Delta t = t_{i+1}-t_{i},$ to smaller values when faster rates are included among the entries in the matrix $\Lambda$. We adopt a similar approach as has been implemented in the Oak Ridge Isotope GENeration (ORIGEN) series of codes \citep{bell1973,croff1980,gauld2011}, where we limit the maximum timestep size $\Delta t$ to an amount inversely proportional to the matrix 1-norm of $\Lambda$, 
\begin{equation}\label{eq:maxdt}
    \Delta t \cdot \|\Lambda \|_1 \leq C \ ,
\end{equation}
for some constant $C.$ For the applications considered in this work, we have found $C\approx 100$ to strike a good balance between numerical accuracy while avoiding the need for prohibitively short timesteps.

We adaptively refine the terms appearing in $\Lambda$ to allow the evolution to progress to longer and longer timescales according to the relationship in Eq.~\ref{eq:maxdt}. In particular, at each timestep $t_i$, we consider the set of nuclear species with nonzero abundances, i.e., the non-zero elements of $\vec{Y}(t_i).$ By traversing all possible decay products that may be produced at later times by only these populated nuclear species, we can eliminate all transitions not associated with this set of possibly-populated nuclear species from the rate matrix. In general, shorter-lived nuclei will completely decay away at early times, so at later times, we only need to track the effects of long-lived nuclei. By restricting the elements of $\Lambda$ to only those associated with these longer-lived nuclei, we can accurately capture their evolution with comparatively longer individual timesteps.

Non-constant decay rates can also be naturally incorporated into the solution for $\vec{Y}(t)$ as might be necessary, for example, when thermally-induced nuclear transitions may arise in high-temperature astrophysical environments \citep{Misch2020} . The strategy implemented in Jade approximates the (time-dependent) decay rates $\lambda$ as constant over each timestep. This imposes a separate upper bound on the maximum timestep size, $\Delta t$, so that all of the $\lambda_{i,j}(t_i)\approx \lambda_{i,j}(t_{i+1}).$ For the applications we consider in this work (the $r$ process) the  transition rates $\lambda$ are non-constant only in the first $\approx 100$ seconds of the evolution, after which they become constant, and the timesteps are no longer subject to this particular constraint. As such, this maximum timestep constraint does not seriously impede our ability to evolve nuclear abundances in these environments over much longer timescales. 

\begin{figure*}
    \centering
    \includegraphics{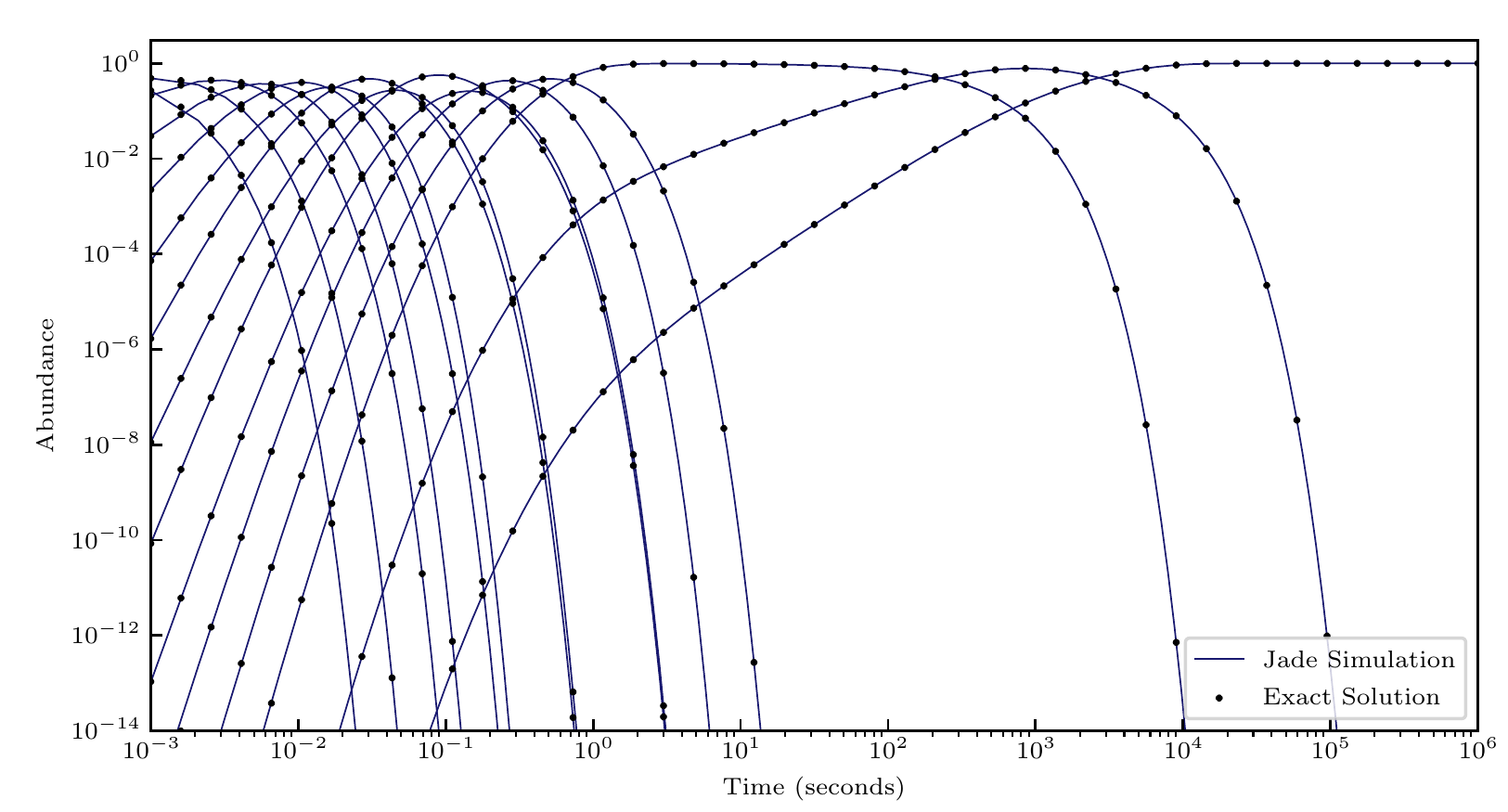}
    \caption{Comparison of Jade decay network solver with exact solution obtained by evaluating Bateman Equations for isotopes along the $A=130$ mass chain, beginning with a unit of abundance located near the neutron drip line at $t=0$. The two approaches are in excellent agreement across a large range of scales in time $(10^{-3}\textrm{ s}<t<10^{6} \textrm{ s})$ and abundance $(10^{-14} < Y < 1)$.}
    \label{fig:benchmark}
\end{figure*}

\subsection{Nuclear Data Considerations}
For the calculations presented in Sec.~\ref{sec:results}, we implement a mixture of experimental, evaluated and theoretical nuclear data. We constrain our simulations to medium-mass nuclei produced immediately following an $r$ process with mass numbers $69 \leq A \leq 204.$ During their subsequent decay towards stable nuclei on much longer timescales, relevant decays are mostly restricted to nuclear $\beta^-$ decays and $\gamma$ transitions between long-lived states of nuclei. Wherever available, we source our $\beta^-$ decay rates from the NuBase (2016) and ENDF-B-VIII.0 compilations \citep{NuBase2016, ENDFB8}. For nuclei where this data is unavailable, we take the decay rates calculated using the Los Alamos Quasi-particle Random Phase Approximation plus Hauser-Feshbach framework \citep{Mumpower2016, Moller2019}.

\subsection{Numerical Accuracy}
We compare the numerical solution to Eq.~\ref{eq:decay_net} obtained with the aforementioned Jade decay network to an exact solution provided by the Bateman Equations. For this test scenario, we consider the evolution of a unit abundance ($Y=1$) of material with mass number $A=130$  located near the neutron dripline, $_{\ 38}^{130}\textrm{Sr}$. We find the two solutions to be in good agreement, with the solution obtained using Jade (blue lines) reproducing the exact solution (black dots) over a range of timescales (milliseconds to days) and abundance ($10^{-14} < Y < 1$). Other isobaric chains express equally good agreement. For the calculations presented in Sec.~\ref{sec:results}, the total solution may be considered as a superposition of many calculations similar to that considered here, and thus we obtain similar accuracy for these `complete' simulations. We restrict ourselves to a single decay chain for the present analysis strictly in the interest of readability of Fig.~\ref{fig:benchmark}.

\begin{figure*}
    \centering
    \includegraphics{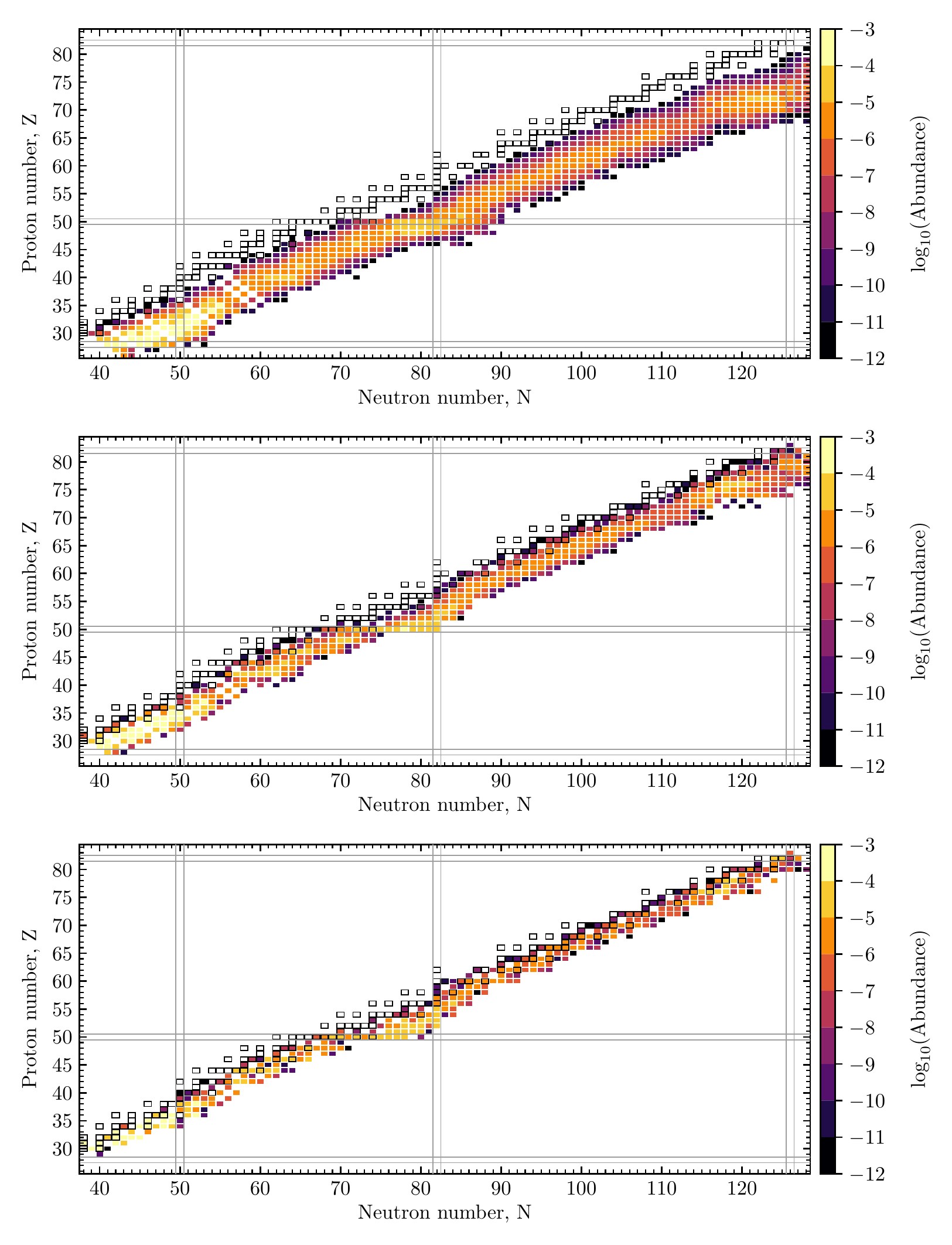}
    \caption{Abundances of individual nuclear species at 1 second (top), 1 minute (middle) and 1 hour (bottom) for an $r$-process-like composition initially situated along the neutron dripline.}
    \label{fig:nz-timesteps}
\end{figure*}

\section{Results} \label{sec:results}
In this work, we focus on the evolution of nuclear abundances in newly-synthesized nuclei in the period immediately following an astrophysical $r$-process event. To begin our analysis, we consider an initial composition placed along the one-neutron dripline as predicted by the FRDM2012 mass table \citep{Moller2016}. The abundance of each nuclear species is then set such that they decay to reproduce the solar isotopic $r$-process residuals reported in \cite{Arnould2007}. We time-evolve these abundances with the Jade decay network over a sequence of timesteps spanning the first 10 years ($\sim 3 \times 10^{8}$ s) of radioactive decay of this system. 

In Fig.~\ref{fig:nz-timesteps}, we plot the abundances of our $r$-process composition at ${1 \textrm{ second}},$ ${1 \textrm{ minute},}$ and ${1 \textrm{ hour}}$ into its evolution. In the first snapshot, at $t=1\textrm{ s},$ the abundances are dispersed over a broad region of the chart of nuclides. However, at successively later times, the abundances become increasingly focused within a narrower range of nuclei \textemdash\ between one and five nuclear species \textemdash\ populated along each isobar at the first hour of evolution.

At least in principle, our ability to time-evolve the decay of $r$-process nuclei should be limited by our knowledge of the distribution of abundance across the many different nuclear species that may be involved in the early-time (dynamic) phase of the $r$-process. However, we show for the first time that a detailed understanding of this early-time evolution is not necessary to confidently simulate the radioactive decay of these nuclei at later times ($t>1\textrm{ minute}$), \textit{provided the final isotopic abundances of the composition are known}. 

To support this assertion, we introduce the idea of \textit{isochronic evolution} of radioactively decaying systems. If we consider a number of short-lived nuclei belonging situated on a common decay series, then they will tend to decay in the short-term in a way that populates a large number of nuclear species with roughly comparable half lives. On longer timescales, however, the decaying system will inevitably populate a species with much longer half life than was present in the initial composition; the entirety of the composition will populate it on a variety of (short) timescales, but the longer half life of this species causes material to saturate its abundance. From this point in the evolution onward, there is no hysteresis regarding the precise distribution of abundance in the initial composition, resulting in an effective `loss of information'. 

\begin{figure*}
    \centering
    \includegraphics{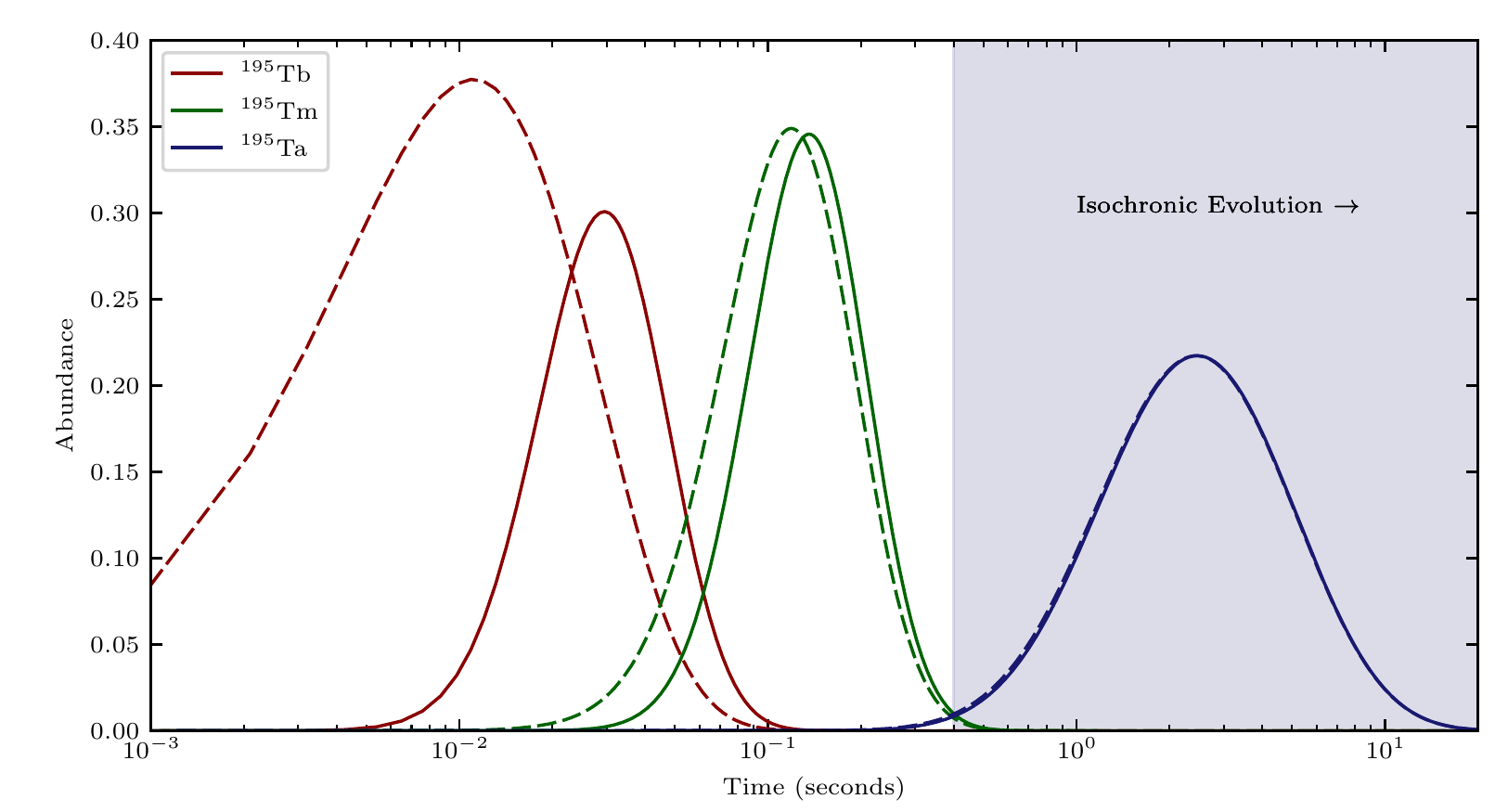}
    \caption{Demonstration of isochronic evolution for a single $(A=195)$ decay chain. Starting composition is at the neutron dripline (dotted) and six units closer to stability (solid). After about a second, the two sets of abundances come into agreement, and their evolution has become isochronic.}
    \label{fig:isochrone-1D}
\end{figure*}

To illustrate this, we consider the decay of nuclei along the $A=195$ mass chain, where we consider two sets of initial conditions. In the first case, we set a unit of abundance on the one-neutron dripline (cerium 195, ${t_{1/2} = 0.6 \textrm{ ms}}$); in the second case, we set a unit of abundance six $\beta^-$ decays closer towards stability (gadolinium 195, ${t_{1/2} = 7.4 \textrm{ ms}}$). For both compositions, we plot in Fig.~\ref{fig:isochrone-1D} the abundances as a function of time for three isotopes along the $A=195$ isobar (terbium 195, ${t_{1/2} = 7.8 \textrm{ ms}}$; thulium 195, ${t_{1/2} = 37 \textrm{ ms}}$; and tantalum 195, ${t_{1/2} = 800 \textrm{ ms}}.$) At early times, $t<1\textrm{ s},$ nuclei with relatively short half lives dominate the composition, and the abundances for the two different initial compositions are in poor agreement. However, at later times ($t > 1 \textrm{ s}$) when comparatively longer-lived nuclei first begin to be populated, both compositions begin to coalesce to a common value, with only nominal relative discrepancies exhibited by the two sets of tantalum 195 abundances. For even longer-lived nuclei that are populated at later times, the two sets of abundances come into even better agreement. We define the \textit{isochronic evolution} of these compositions to be that period of time in which abundances tend to be similar, regardless of the specific arrangement of abundance across their shorter-lived progenitors. 

To explore the robustness of the isochronic evolution in $r$ process compositions, we explore the temporal evolution of nuclear abundances for one million randomly selected nuclear compositions that simultaneously (1) are representative of a composition shortly following neutron exhaustion in an r process and (2) are constrained to reproduce the solar isotopic r process abundances of \cite{Arnould2007}. To achieve this, we constrain ourselves to compositions that include only the short-lived, neutron-rich nuclei that the $r$ process proceeds through, where we set our cutoff at nuclei with half-lives shorter than 1 second. We also constrain our compositions such that all nuclei are one- and two-neutron bound, i.e., our samples lie entirely within the neutron dripline. Within these bounds, we use a random number generator to set the abundances along each isobar, where an overall scale factor is applied to the random numbers such that the summed abundance along the isobar reproduces the solar abundance attributed to the same isobar.

We process each of our one million samples through the Jade decay network solver to simulate each abundance evolution. To assist the analysis and discussion of these samples, we define the function 
\begin{equation}\label{eq:uncert}
    \delta Y(i,j;t) = \max{Y_{i,j}(t)}-\min{Y_{i,j}(t)} \ ,
\end{equation}
where $\max{Y_{i,j}(t)}$ and $\min{Y_{i,j}(t)}$ give the maximum and minimum abundance of species $(i,j)$ at time $t$ across all samples, and $\delta Y(i,j;t)$ gives an upper bound on the difference between any two samples. The quantity $\delta Y(i,j;t)$ may be intuitively understood as the uncertainty in simulated abundances due to uncertainty in the exact distribution of an $r$-process-like composition. We also define the related quantity
\begin{equation}\label{eq:uncert_norm}
    \overline{\delta Y}(i,j;t) = \frac{\delta Y(i,j;t)} { \sqrt{\max{Y_{i,j}(t)} \times \min{Y_{i,j}}(t)} } \ ,
\end{equation}
where $\overline{\delta Y}$ normalizes $\delta Y$ to the geometric mean of the upper and lower bounds of the sampled abundances, thereby expressing $\delta Y$ relative to a `typical' value of the respective abundance.

\begin{figure*}
    \centering
    \includegraphics{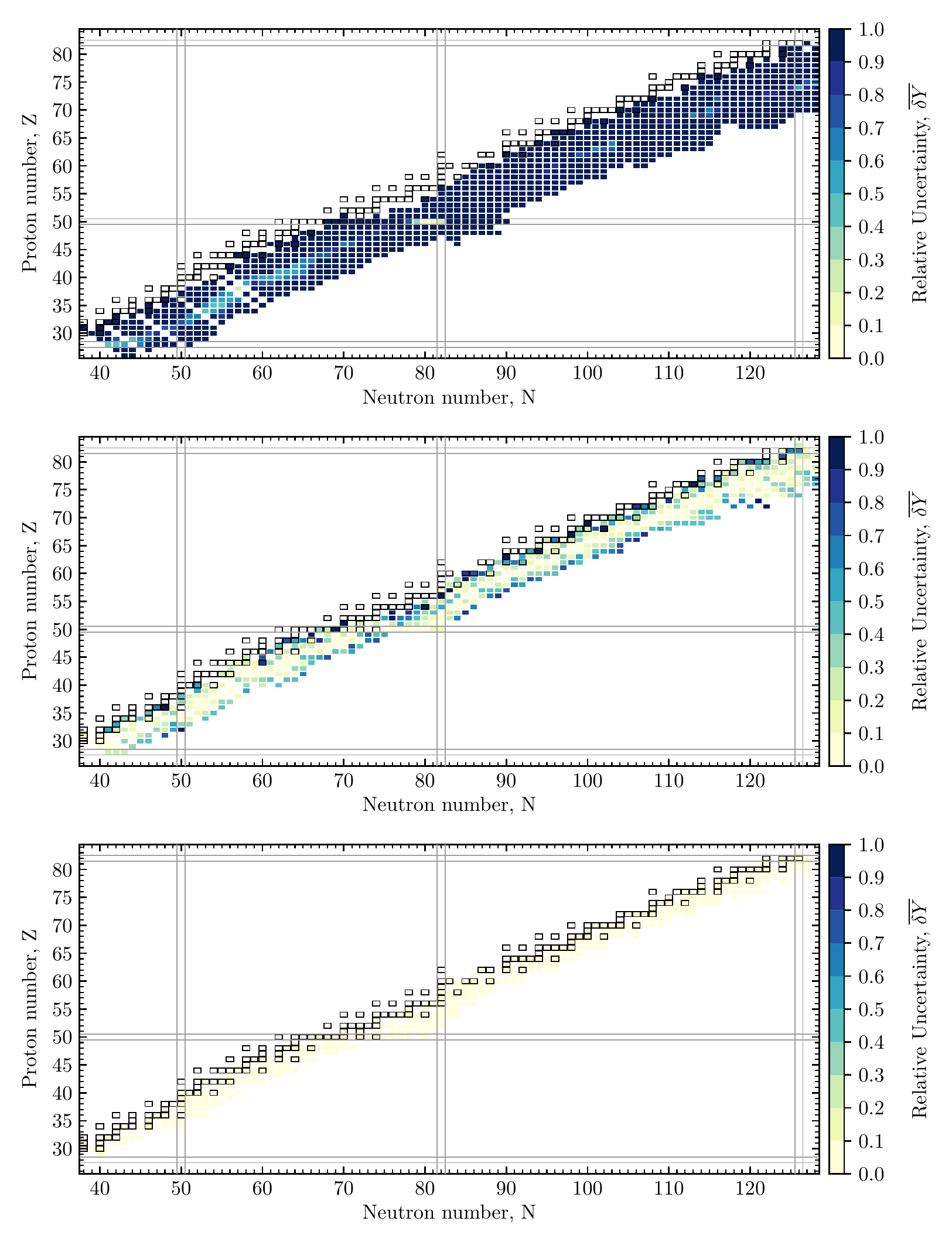}
    \caption{Relative abundance variations ($\overline{\delta Y},$ as defined by Eq.~\ref{eq:uncert_norm}.) from ${1,000,000}$ samples of $r$-process-like nuclear compositions at 1 s (top), 1 min (middle), and 1 hour (bottom).}
    \label{fig:rp-uncert-NZ}
\end{figure*}

In Fig.~\ref{fig:rp-uncert-NZ}, we plot $\overline{\delta Y}$ at the same three timesteps as was used in Fig.~\ref{fig:nz-timesteps}, i.e., 1 second, 1 minute, and 1 hour. At 1 second, the value of $\overline{\delta Y}$ is nearly or well in excess of 1, indicating that the range in abundances across the chart of nuclides is highly uncertain, as many nuclei are populated whose half lives are still comparable to the half lives characterising the initial composition.

Within 1 minute into the evolution, however, we find that all of our compositions have begun to transition to an isochronic phase of their evolution \textemdash\ the most abundant components at this point in time have a typical relative uncertainty $\overline{\delta Y} <  10\%$ \textemdash\ with the larger relative uncertainties generally present in the less-abundant tails of the abundance pattern.

Finally, by 1 hour (and all later points in time), all of our million samples have fully transitioned into an isochronic evolution, with all abundances at this time in agreement to within $\overline{\delta Y }< 2.5\%$, and most abundances in agreement to well within $\overline{\delta Y }\ll 1\%.$ 

\begin{figure}
    \includegraphics{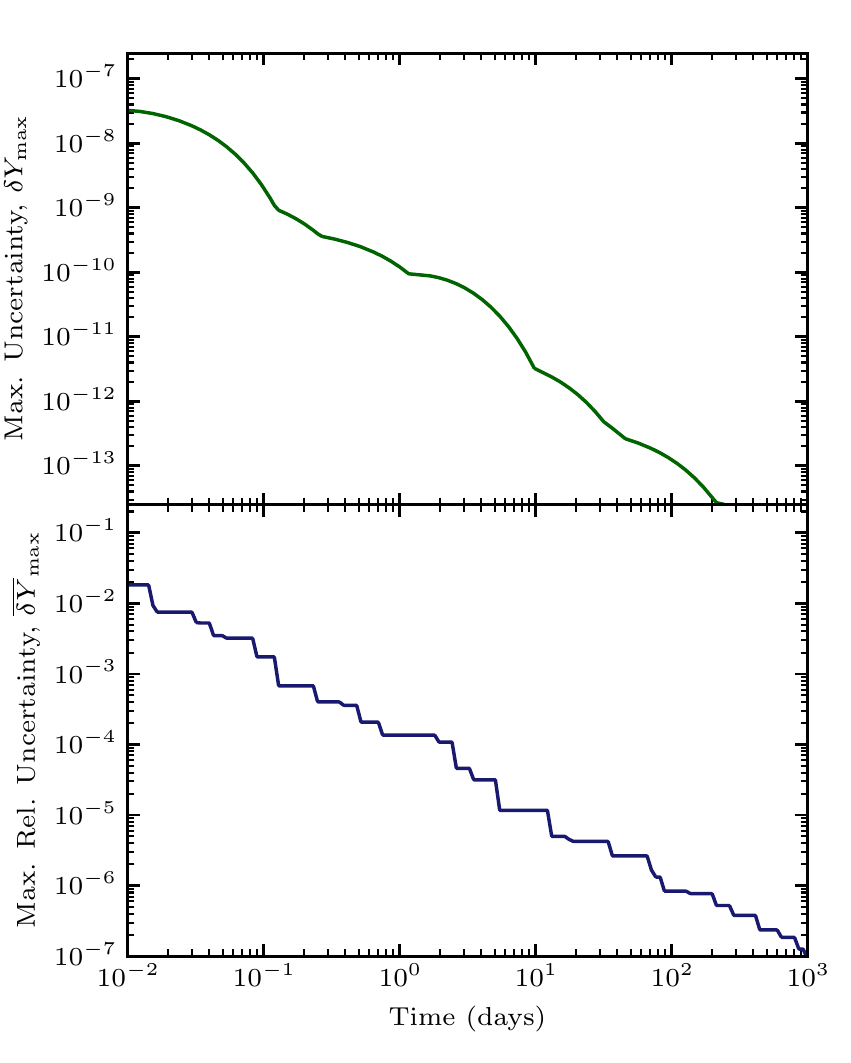}
    \caption{Upper bounds on uncertainty for decay network simulations of ${1,000,000}$ samples of $r$-process-like nuclear compositions. Top: Maximum range in abundance across all nuclear species as a function of time ($\delta Y_{\text{max}}$ as defined in Eq.~\ref{eq:uncert_max}). Bottom: Maximum relative error in abundance across all nuclear species as a function of time ($\overline{\delta Y}_{\text{max}}$, as defined in Eq.~\ref{eq:uncert_relmax}). By 0.01 day, \textit{all} simulated abundances come into agreement to within $\sim 2.5\%$ and continue to decrease according to  a power-law relationship in time.}
    \label{fig:rp-uncert-1D}
\end{figure}

As a final reinforcement of the isochronic nature of our simulated $r$-process decay systems, in Fig.~\ref{fig:rp-uncert-1D} we plot as a function of time two additional metrics that summarize Eq.~\ref{eq:uncert} and Eq.~\ref{eq:uncert_norm} across all nuclear species. In the first case, we define
\begin{equation}\label{eq:uncert_max}
    \delta Y_{\text{max}}(t) = \max{\left\{\delta Y(i,j;t)\right\}}\ ,
\end{equation}
which  gives the maximum value of $\delta Y$ across all nuclear species in the decay network, giving an upper bound on the total uncertainty of any one abundance. The upper panel of Fig.~\ref{fig:rp-uncert-1D} plots the evolution of $\delta Y_{\text{max}}$ over time for our abundance calculations. The species with the largest values of $\delta Y$ tend to also be the more abundant $(Y \gtrsim 10^-4),$ such that they maximize $\delta Y$ despite having the lowest relative errors, $\overline{\delta Y},$ overall. In this sense, $\delta Y_{\text{max}}$ provides a representative upper bound on abundance variations for the most abundant nuclei.      

Similarly, we define the analogous quantity for $\overline{\delta Y},$
\begin{equation}\label{eq:uncert_relmax}
   \overline{\delta Y}_{\text{max}}(t) = \max{\left\{\overline{\delta Y}(i,j;t) \right\}}\ ,
\end{equation}
which we plot in the lower panel of Fig.~\ref{fig:rp-uncert-1D} and where we additionally impose an abundance cutoff of ${ Y > 10^{-14}}$ to avoid large \textit{relative} uncertainties in otherwise-negligible abundances. In contrast with $\delta Y_{\text{max}},$ the species which define $\overline{\delta Y}_{\text{max}}$ by maximizing $\overline{\delta Y}$ are also the least abundant, leading to the stair-step behavior exhibited in Fig.~\ref{fig:rp-uncert-1D}. The sudden drops in $\overline{\delta Y}_{\text{max}}$ correspond to whichever species maximizes $\overline{\delta Y}$ dropping below our $Y>10^{-14}$ threshold, causing a discrete jump to the next-largest value of $\delta Y$ among species populated in the system. Consequently, $\overline{\delta Y}_{\text{max}}$ is a representative upper bound on abundance variations in the least abundant nuclei.

In any case, we note that by 15 minutes into the decay of our compositions, all of our million $r$-process composition samples agree to within $1\%,$ and certainly by $1 \textrm{ day}$ all abundances agree to within one part in ten thousand. As such, we conclude that the decay-phase of the $r$ process is a strong example of the isochronic evolution we have introduced in this work. This allows us to describe late-time (${t\gtrsim 15 \textrm{ min}}$) behavior of $r$-process ejecta based on the assumptions that
\begin{enumerate}
    \item all nuclei produced during the $r$ process are sufficiently neutron-rich that their half-lives are shorter than 1 second, and 
    \item the final isotopic abundances of nuclei produced during an r-process event are known.
\end{enumerate}
Provided we admit these assumptions, this enables us to reliably predict the evolution of nuclear abundances and the radioactive decay that follows from $r$-process nucleosynthesis while avoiding a number of difficult issues pertaining to the large uncertainties in the physical properties of the most neutron-rich nuclei that directly affect predictions based on direct modeling of nucleosynthesis, e.g., \cite{Grossman2014, Martin2016, Lippuner2017, Fernandex2017, Radice2018, Wollaeger2018, Zhu2018, Miller2019, Miller2020, Barnes2020, Wesley2020, Zhu2020}. As such, nucleosynthesis calculations performed via the method outlined in this work provide an alternate, but complimentary, approach to understanding the late-time radioactive decay of freshly synthesized $r$-process nuclei. 

In the Appendix, we provide results of a first calculation based on our approach as a series of abundance tables for  timesteps ranging from 15 minutes to 10 years following a `full' $r$ process producing the first, second, and third $r$-process peaks. These abundances, along with their associated decay properties, may be directly implemented into calculations simulating late-time observables associated with candidate $r$-process sites, as well as in the physical interpretation of existing kilonova observations, e.g., the binary neutron star merger observations of GW170817/AT2017gfo/GRB170817A \citep{Metzger2010,Metzger2012,Metzger2017,Metzger2019,Cowperthwaite2017,Tanvir2017}. 

We note that our procedure can readily be applied to investigate a wide range of potential $r$-process scenarios, for example, an astrophysical event producing the first ($A\sim 80$) $r$-process peak in isolation. We restrict our calculations to an $r$-process event producing the first three peaks in the solar isotopic ratios in this study. Furthermore, while we do not address it in the present work, the late-time decay of heavier species, e.g. the production and subsequent decay of long-lived actinides \citep{Korobkin2012, Wanajo2014, Eichler2015, Zhu2018, Holmbeck2019, Holmbeck2019b, Holmbeck2020}, may also be studied in our approach. However, the intricate mixture of nuclear $\alpha$ decay, $\beta$ decay, and various modes of fission experienced by the heaviest nuclei that may be produced during an $r$ process reintroduces the nuclear physics-associated uncertainties in a way that is not easily resolved using the techniques employed in this work \citep{Vassh2019,Wu2019,Barnes2020,Vassh2020,Zhu2020}. Furthermore, the lack of stable nuclei with mass numbers $A>209$ introduces substantial degeneracy to the problem of choosing and/or sampling from the space of reasonable initial compositions for the heaviest nuclei, insofar as one must constrain the abundances of hundreds of the heaviest nuclei using the abundances of only a half dozen or so stable or extremely long-lived ($t_{1/2}\gtrsim 100~\textrm{Myr}$) isotopes in which their decay series terminate, fission products notwithstanding. In future work, we will address these challenges directly, in particular by systematically probing the unique uncertainties discussed here, and propagating their combined effects to simulated $r$-process nucleosynthesis and corresponding electromagnetic observables.

\section{Conclusions} \label{sec:concl}
In this work, we have explored the radioactive decay of medium-mass, neutron-rich nuclei in the period of time following their synthesis via rapid neutron capture. In this epoch of nucleosynthesis, free neutrons are all but completely exhausted, and nuclear decay processes control the subsequent evolution of the nuclear abundances. In the absence of two- or more-body reactions between nuclei, the standard nuclear reaction network equations assume a simplified form. We have introduced a new nuclear decay network, Jade, and have discussed several aspects of its implementation. 

Using Jade, we have simulated a large number of nuclear systems representative of those present in the immediate aftermath of an $r$ process. From these calculations, we have highlighted the tendency of the sampled compositions to converge to a single effective evolution, which we attribute to steep gradients in the decay rates encountered as nuclei decay towards stability. We have introduced the term `isochronic evolution' to describe this general property of decay systems progressing from short-lived initial states through longer lived intermediate and final states. While we have not explored them in this work, the techniques we have employed may reasonably be adapted for the description and analysis of a broad range of physical applications governed by qualitatively similar systems of ODEs. Of particular interest would be its application to the study of the de-excitation and decay of fission fragments, as well as its application to the study of isotope generation within reactor environments.

During the isochronic phase of decaying $r$-process compositions, the populated nuclei are generally well-studied experimentally (including their atomic masses and decay half lives). As such, we can simulate their radioactive decay and relative abundances with a relatively high degree of confidence, with the primary source of uncertainty in this approach being the final distribution of nuclear abundances. Our approach compliments existing methods for predicting nucleosynthesis in $r$-process environments based on direct modeling. In order to facilitate future studies, we have provided in the Appendix as supplementary material snapshots of nuclear abundances obtained from our calculations for $r$ process compositions reproducing solar $r$-process residuals in the mass range $69 \leq A \leq 204$.

\section{Acknowledgements}\label{sec:ack}
T.M.S., G.W.M and M.R.M. were supported by the US Department of Energy through the Los Alamos National Laboratory (LANL). LANL is operated by Triad National Security, LLC, for the National Nuclear Security Administration of U.S.\ Department of Energy (Contract No.\ 89233218CNA000001). 
T.M.S. was partly supported by the Fission In R-process Elements (FIRE) Topical Collaboration in Nuclear Theory, funded by the U.S. Department of Energy. 
G.W.M and M.R.M. were partly supported by the Laboratory Directed Research and Development program of LANL under project number 20190021DR. 

\software{
MatPlotLib \citep[v3.3.1; ][]{Hunter07},
NumPy \citep[v1.19.0; ][]{VanDerWalt2011},
SciPy \citep[v1.5.2; ][]{AlMohy2010}
}

\bibliography{refs}{}

\begin{thebibliography}{}
\expandafter\ifx\csname natexlab\endcsname\relax\def\natexlab#1{#1}\fi
\providecommand{\url}[1]{\href{#1}{#1}}
\providecommand{\dodoi}[1]{doi:~\href{http://doi.org/#1}{\nolinkurl{#1}}}
\providecommand{\doeprint}[1]{\href{http://ascl.net/#1}{\nolinkurl{http://ascl.net/#1}}}
\providecommand{\doarXiv}[1]{\href{https://arxiv.org/abs/#1}{\nolinkurl{https://arxiv.org/abs/#1}}}

\bibitem[{{Abbott} {et~al.}(2017){Abbott}, {Abbott}, {Abbott}, {Acernese},
  {Ackley}, {Adams}, {Adams}, {Addesso}, {Adhikari}, {Adya}, \&
  et~al.}]{Abbott2017a}
{Abbott}, B.~P., {Abbott}, R., {Abbott}, T.~D., {et~al.} 2017, \apjl, 848, L12,
  \dodoi{10.3847/2041-8213/aa91c9}

\bibitem[{Al-Mohy \& Higham(2010)}]{AlMohy2010}
Al-Mohy, A.~H., \& Higham, N.~J. 2010, SIAM Journal on Matrix Analysis and
  Applications, 31, 970, \dodoi{10.1137/09074721X}

\bibitem[{{Aprahamian} {et~al.}(2014){Aprahamian}, {Bentley}, {Mumpower}, \&
  {Surman}}]{Aprahamian2014}
{Aprahamian}, A., {Bentley}, I., {Mumpower}, M., \& {Surman}, R. 2014, AIP
  Advances, 4, 041101, \dodoi{10.1063/1.4867193}

\bibitem[{{Arnould} \& {Goriely}(2020)}]{Arnould2020}
{Arnould}, M., \& {Goriely}, S. 2020, Progress in Particle and Nuclear Physics,
  112, 103766, \dodoi{10.1016/j.ppnp.2020.103766}

\bibitem[{{Arnould} {et~al.}(2007){Arnould}, {Goriely}, \&
  {Takahashi}}]{Arnould2007}
{Arnould}, M., {Goriely}, S., \& {Takahashi}, K. 2007, \physrep, 450, 97,
  \dodoi{10.1016/j.physrep.2007.06.002}

\bibitem[{Audi {et~al.}(2017)Audi, Kondev, Wang, Huang, \& Naimi}]{NuBase2016}
Audi, G., Kondev, F.~G., Wang, M., Huang, W.~J., \& Naimi, S. 2017, Chin. Phys.
  C, 41, 030001, \dodoi{10.1088/1674-1137/41/3/030001}

\bibitem[{{Baldo} {et~al.}(2017){Baldo}, {Robledo}, {Schuck}, \&
  {Vi{\~n}as}}]{Baldo2017}
{Baldo}, M., {Robledo}, L.~M., {Schuck}, P., \& {Vi{\~n}as}, X. 2017, \prc, 95,
  014318, \dodoi{10.1103/PhysRevC.95.014318}

\bibitem[{{Barnes} {et~al.}(2020){Barnes}, {Zhu}, {Lund}, {Sprouse}, {Vassh},
  {McLaughlin}, {Mumpower}, \& {Surman}}]{Barnes2020}
{Barnes}, J., {Zhu}, Y.~L., {Lund}, K.~A., {et~al.} 2020, arXiv e-prints,
  arXiv:2010.11182.
\newblock \doarXiv{2010.11182}

\bibitem[{Bateman(1908)}]{bateman1910}
Bateman, H. 1908, Proc. Cambridge Phil. Soc., 1908, 15, 423

\bibitem[{Bell(1973)}]{bell1973}
Bell, M. 1973, ORIGEN: the ORNL isotope generation and depletion code, Tech.
  rep., Oak Ridge National Lab., Tenn.(USA)

\bibitem[{{Bliss} {et~al.}(2017){Bliss}, {Arcones}, {Montes}, \&
  {Pereira}}]{Bliss2017}
{Bliss}, J., {Arcones}, A., {Montes}, F., \& {Pereira}, J. 2017, Journal of
  Physics G Nuclear Physics, 44, 054003, \dodoi{10.1088/1361-6471/aa63bd}

\bibitem[{Brown {et~al.}(2018)Brown, Chadwick, Capote, Kahler, Trkov, Herman,
  Sonzogni, Danon, Carlson, Dunn, Smith, Hale, Arbanas, Arcilla, Bates, Beck,
  Becker, Brown, Casperson, Conlin, Cullen, Descalle, Firestone, Gaines, Guber,
  Hawari, Holmes, Johnson, Kawano, Kiedrowski, Koning, Kopecky, Leal, Lestone,
  Lubitz, {Márquez Damián}, Mattoon, McCutchan, Mughabghab, Navratil,
  Neudecker, Nobre, Noguere, Paris, Pigni, Plompen, Pritychenko, Pronyaev,
  Roubtsov, Rochman, Romano, Schillebeeckx, Simakov, Sin, Sirakov, Sleaford,
  Sobes, Soukhovitskii, Stetcu, Talou, Thompson, {van der Marck},
  Welser-Sherrill, Wiarda, White, Wormald, Wright, Zerkle, Žerovnik, \&
  Zhu}]{ENDFB8}
Brown, D., Chadwick, M., Capote, R., {et~al.} 2018, Nuclear Data Sheets, 148, 1
  , \dodoi{https://doi.org/10.1016/j.nds.2018.02.001}

\bibitem[{{Bulgac} {et~al.}(2018){Bulgac}, {Jin}, {Roche}, {Schunck}, \&
  {Stetcu}}]{Bulgac+18}
{Bulgac}, A., {Jin}, S., {Roche}, K., {Schunck}, N., \& {Stetcu}, I. 2018,
  arXiv e-prints.
\newblock \doarXiv{1806.00694}

\bibitem[{Burbidge {et~al.}(1957)Burbidge, Burbidge, Fowler, \&
  Hoyle}]{Burbidge1957}
Burbidge, E.~M., Burbidge, G.~R., Fowler, W.~A., \& Hoyle, F. 1957, Rev. Mod.
  Phys., 29, 547, \dodoi{10.1103/RevModPhys.29.547}

\bibitem[{{Cameron}(1957)}]{Cameron1957}
{Cameron}, A.~G.~W. 1957, \pasp, 69, 201, \dodoi{10.1086/127051}

\bibitem[{Cetnar(2006)}]{cetnar2006}
Cetnar, J. 2006, Annals of Nuclear Energy, 33, 640

\bibitem[{{Coc} {et~al.}(2000){Coc}, {Porquet}, \& {Nowacki}}]{Coc2000}
{Coc}, A., {Porquet}, M.-G., \& {Nowacki}, F. 2000, \prc, 61, 015801,
  \dodoi{10.1103/PhysRevC.61.015801}

\bibitem[{{Cowan} {et~al.}(2019){Cowan}, {Sneden}, {Lawler}, {Aprahamian},
  {Wiescher}, {Langanke}, {Mart{\'\i}nez-Pinedo}, \& {Thielemann}}]{Cowan2019}
{Cowan}, J.~J., {Sneden}, C., {Lawler}, J.~E., {et~al.} 2019, arXiv e-prints,
  arXiv:1901.01410.
\newblock \doarXiv{1901.01410}

\bibitem[{{Cowperthwaite \textit{et al.}}(2017)}]{Cowperthwaite2017}
{Cowperthwaite \textit{et al.}}, P.~S. 2017, The Astrophysical Journal Letters,
  848, L17.
\newblock \url{http://stacks.iop.org/2041-8205/848/i=2/a=L17}

\bibitem[{Croff(1980)}]{croff1980}
Croff, A.~G. 1980, User's manual for the ORIGEN2 computer code, Tech. rep., Oak
  Ridge National Lab.

\bibitem[{{Cuenca-Garc{\i}a} {et~al.}(2007){Cuenca-Garc{\i}a},
  {Mart{\i}nez-Pinedo}, {Langanke}, {Nowacki}, \& {Borzov}}]{Cuenca2007}
{Cuenca-Garc{\i}a}, J.~J., {Mart{\i}nez-Pinedo}, G., {Langanke}, K., {Nowacki},
  F., \& {Borzov}, I.~N. 2007, European Physical Journal A, 34, 99,
  \dodoi{10.1140/epja/i2007-10477-3}

\bibitem[{Dillmann \& Tarifeño-Saldivia(2018)}]{Dillmann2018}
Dillmann, I., \& Tarifeño-Saldivia, A. 2018, Nuclear Physics News, 28, 28,
  \dodoi{10.1080/10619127.2018.1427937}

\bibitem[{{Eichler} {et~al.}(2015){Eichler}, {Arcones}, {Kelic}, {Korobkin},
  {Langanke}, {Marketin}, {Martinez-Pinedo}, {Panov}, {Rauscher}, {Rosswog},
  {Winteler}, {Zinner}, \& {Thielemann}}]{Eichler2015}
{Eichler}, M., {Arcones}, A., {Kelic}, A., {et~al.} 2015, \apj, 808, 30,
  \dodoi{10.1088/0004-637X/808/1/30}

\bibitem[{{Erler} {et~al.}(2012){Erler}, {Birge}, {Kortelainen}, {Nazarewicz},
  {Olsen}, {Perhac}, \& {Stoitsov}}]{Erler2012}
{Erler}, J., {Birge}, N., {Kortelainen}, M., {et~al.} 2012, \nat, 486, 509,
  \dodoi{10.1038/nature11188}

\bibitem[{{Even} {et~al.}(2020){Even}, {Korobkin}, {Fryer}, {Fontes},
  {Wollaeger}, {Hungerford}, {Lippuner}, {Miller}, {Mumpower}, \&
  {Misch}}]{Wesley2020}
{Even}, W., {Korobkin}, O., {Fryer}, C.~L., {et~al.} 2020, \apj, 899, 24,
  \dodoi{10.3847/1538-4357/ab70b9}

\bibitem[{{Fern{\'a}ndez} {et~al.}(2017){Fern{\'a}ndez}, {Foucart}, {Kasen},
  {Lippuner}, {Desai}, \& {Roberts}}]{Fernandex2017}
{Fern{\'a}ndez}, R., {Foucart}, F., {Kasen}, D., {et~al.} 2017, Classical and
  Quantum Gravity, 34, 154001, \dodoi{10.1088/1361-6382/aa7a77}

\bibitem[{{Freiburghaus} {et~al.}(1999){Freiburghaus}, {Rosswog}, \&
  {Thielemann}}]{Freiburghaus1999}
{Freiburghaus}, C., {Rosswog}, S., \& {Thielemann}, F.~K. 1999, \apjl, 525,
  L121, \dodoi{10.1086/312343}

\bibitem[{Furuta {et~al.}(1987)Furuta, Oka, \& Kondo}]{furuta1987}
Furuta, K., Oka, Y., \& Kondo, S. 1987, CCC-464) Informal Notes (February 1987)

\bibitem[{Gauld {et~al.}(2011)Gauld, Radulescu, Ilas, Murphy, Williams, \&
  Wiarda}]{gauld2011}
Gauld, I.~C., Radulescu, G., Ilas, G., {et~al.} 2011, Nuclear Technology, 174,
  169, \dodoi{10.13182/NT11-3}

\bibitem[{Giuliani {et~al.}(2018)Giuliani, Mart\'{\i}nez-Pinedo, \&
  Robledo}]{Giuliani2018}
Giuliani, S.~A., Mart\'{\i}nez-Pinedo, G., \& Robledo, L.~M. 2018, Phys. Rev.
  C, 97, 034323, \dodoi{10.1103/PhysRevC.97.034323}

\bibitem[{{Giuliani} {et~al.}(2020){Giuliani}, {Mart{\'\i}nez-Pinedo}, {Wu}, \&
  {Robledo}}]{Giuliani2020}
{Giuliani}, S.~A., {Mart{\'\i}nez-Pinedo}, G., {Wu}, M.-R., \& {Robledo}, L.~M.
  2020, \prc, 102, 045804, \dodoi{10.1103/PhysRevC.102.045804}

\bibitem[{{Goriely} \& {Mart{\'\i}nez Pinedo}(2015)}]{Goriely2015}
{Goriely}, S., \& {Mart{\'\i}nez Pinedo}, G. 2015, \nphysa, 944, 158,
  \dodoi{10.1016/j.nuclphysa.2015.07.020}

\bibitem[{{Grossman} {et~al.}(2014){Grossman}, {Korobkin}, {Rosswog}, \&
  {Piran}}]{Grossman2014}
{Grossman}, D., {Korobkin}, O., {Rosswog}, S., \& {Piran}, T. 2014, \mnras,
  439, 757, \dodoi{10.1093/mnras/stt2503}

\bibitem[{{Hix} \& {Thielemann}(1999)}]{Hix1999}
{Hix}, W.~R., \& {Thielemann}, F.~K. 1999, Journal of Computational and Applied
  Mathematics, 109, 321.
\newblock \doarXiv{astro-ph/9906478}

\bibitem[{{Holmbeck} {et~al.}(2019{\natexlab{a}}){Holmbeck}, {Frebel},
  {McLaughlin}, {Mumpower}, {Sprouse}, \& {Surman}}]{Holmbeck2019}
{Holmbeck}, E.~M., {Frebel}, A., {McLaughlin}, G.~C., {et~al.}
  2019{\natexlab{a}}, \apj, 881, 5, \dodoi{10.3847/1538-4357/ab2a01}

\bibitem[{{Holmbeck} {et~al.}(2019{\natexlab{b}}){Holmbeck}, {Frebel},
  {McLaughlin}, {Mumpower}, {Sprouse}, \& {Surman}}]{Holmbeck2019b}
---. 2019{\natexlab{b}}, \apj, 881, 5, \dodoi{10.3847/1538-4357/ab2a01}

\bibitem[{{Holmbeck} {et~al.}(2020){Holmbeck}, {Frebel}, {McLaughlin},
  {Surman}, {Fernandez}, {Metzger}, {Mumpower}, \& {Sprouse}}]{Holmbeck2020}
---. 2020, arXiv e-prints, arXiv:2010.01621.
\newblock \doarXiv{2010.01621}

\bibitem[{{Horowitz} {et~al.}(2019){Horowitz}, {Arcones}, {C{\^o}t{\'e}},
  {Dillmann}, {Nazarewicz}, {Roederer}, {Schatz}, {Aprahamian}, {Atanasov},
  {Bauswein}, {Beers}, {Bliss}, {Brodeur}, {Clark}, {Frebel}, {Foucart},
  {Hansen}, {Just}, {Kankainen}, {McLaughlin}, {Kelly}, {Liddick}, {Lee},
  {Lippuner}, {Martin}, {Mendoza-Temis}, {Metzger}, {Mumpower}, {Perdikakis},
  {Pereira}, {O'Shea}, {Reifarth}, {Rogers}, {Siegel}, {Spyrou}, {Surman},
  {Tang}, {Uesaka}, \& {Wang}}]{Horowitz2019}
{Horowitz}, C.~J., {Arcones}, A., {C{\^o}t{\'e}}, B., {et~al.} 2019, Journal of
  Physics G Nuclear Physics, 46, 083001, \dodoi{10.1088/1361-6471/ab0849}

\bibitem[{{Hosmer} {et~al.}(2005){Hosmer}, {Schatz}, {Aprahamian}, {Arndt},
  {Clement}, {Estrade}, {Kratz}, {Liddick}, {Mantica}, {Mueller}, {Montes},
  {Morton}, {Ouellette}, {Pellegrini}, {Pfeiffer}, {Reeder}, {Santi},
  {Steiner}, {Stolz}, {Tomlin}, {Walters}, \& {W{\"o}hr}}]{Hosmer2005}
{Hosmer}, P.~T., {Schatz}, H., {Aprahamian}, A., {et~al.} 2005, \prl, 94,
  112501, \dodoi{10.1103/PhysRevLett.94.112501}

\bibitem[{{Hunter}(2007)}]{Hunter07}
{Hunter}, J.~D. 2007, Computing in Science Engineering, 9, 90,
  \dodoi{10.1109/MCSE.2007.55}

\bibitem[{{Kajino} {et~al.}(2019){Kajino}, {Aoki}, {Balantekin}, {Diehl},
  {Famiano}, \& {Mathews}}]{Kajino2019}
{Kajino}, T., {Aoki}, W., {Balantekin}, A.~B., {et~al.} 2019, Progress in
  Particle and Nuclear Physics, 107, 109, \dodoi{10.1016/j.ppnp.2019.02.008}

\bibitem[{{Korobkin} {et~al.}(2012){Korobkin}, {Rosswog}, {Arcones}, \&
  {Winteler}}]{Korobkin2012}
{Korobkin}, O., {Rosswog}, S., {Arcones}, A., \& {Winteler}, C. 2012, \mnras,
  426, 1940, \dodoi{10.1111/j.1365-2966.2012.21859.x}

\bibitem[{{Korobkin} {et~al.}(2020){Korobkin}, {Hungerford}, {Fryer},
  {Mumpower}, {Misch}, {Sprouse}, {Lippuner}, {Surman}, {Couture}, {Bloser},
  {Shirazi}, {Even}, {Vestrand}, \& {Miller}}]{Korobkin2020}
{Korobkin}, O., {Hungerford}, A.~M., {Fryer}, C.~L., {et~al.} 2020, \apj, 889,
  168, \dodoi{10.3847/1538-4357/ab64d8}

\bibitem[{{Li} \& {Paczy{\'n}ski}(1998)}]{Li1998}
{Li}, L.-X., \& {Paczy{\'n}ski}, B. 1998, \apjl, 507, L59,
  \dodoi{10.1086/311680}

\bibitem[{{Liddick} {et~al.}(2016){Liddick}, {Spyrou}, {Crider}, {Naqvi},
  {Larsen}, {Guttormsen}, {Mumpower}, {Surman}, {Perdikakis}, {Bleuel},
  {Couture}, {Crespo Campo}, {Dombos}, {Lewis}, {Mosby}, {Nikas}, {Prokop},
  {Renstrom}, {Rubio}, {Siem}, \& {Quinn}}]{Liddick2016}
{Liddick}, S.~N., {Spyrou}, A., {Crider}, B.~P., {et~al.} 2016, \prl, 116,
  242502, \dodoi{10.1103/PhysRevLett.116.242502}

\bibitem[{{Lippuner} \& {Roberts}(2017)}]{Lippuner2017}
{Lippuner}, J., \& {Roberts}, L.~F. 2017, The Astrophysical Journal Supplement
  Series, 233, 18, \dodoi{10.3847/1538-4365/aa94cb}

\bibitem[{{Lyons} {et~al.}(2019){Lyons}, {Spyrou}, {Liddick}, {Naqvi},
  {Crider}, {Dombos}, {Bleuel}, {Brown}, {Couture}, {Campo}, {Engel},
  {Guttormsen}, {Larsen}, {Lewis}, {M{\"o}ller}, {Mosby}, {Mumpower}, {Ney},
  {Palmisano}, {Perdikakis}, {Prokop}, {Renstr{\o}m}, {Siem}, {Smith}, \&
  {Quinn}}]{Lyons2019}
{Lyons}, S., {Spyrou}, A., {Liddick}, S.~N., {et~al.} 2019, \prc, 100, 025806,
  \dodoi{10.1103/PhysRevC.100.025806}

\bibitem[{{Marketin} {et~al.}(2016){Marketin}, {Huther}, \&
  {Mart{\'\i}nez-Pinedo}}]{Marketin2016}
{Marketin}, T., {Huther}, L., \& {Mart{\'\i}nez-Pinedo}, G. 2016, \prc, 93,
  025805, \dodoi{10.1103/PhysRevC.93.025805}

\bibitem[{{Martin} {et~al.}(2016){Martin}, {Arcones}, {Nazarewicz}, \&
  {Olsen}}]{Martin2016}
{Martin}, D., {Arcones}, A., {Nazarewicz}, W., \& {Olsen}, E. 2016, \prl, 116,
  121101, \dodoi{10.1103/PhysRevLett.116.121101}

\bibitem[{Mart\'{\i}nez-Pinedo \& Langanke(1999)}]{Pinedo1999}
Mart\'{\i}nez-Pinedo, G., \& Langanke, K. 1999, Phys. Rev. Lett., 83, 4502,
  \dodoi{10.1103/PhysRevLett.83.4502}

\bibitem[{{Mathews} \& {Cowan}(1990)}]{Mathews1990}
{Mathews}, G.~J., \& {Cowan}, J.~J. 1990, \nat, 345, 491,
  \dodoi{10.1038/345491a0}

\bibitem[{{McDonnell} {et~al.}(2015){McDonnell}, {Schunck}, {Higdon}, {Sarich},
  {Wild}, \& {Nazarewicz}}]{McDonnell2015}
{McDonnell}, J.~D., {Schunck}, N., {Higdon}, D., {et~al.} 2015, \prl, 114,
  122501, \dodoi{10.1103/PhysRevLett.114.122501}

\bibitem[{{Metzger}(2017)}]{Metzger2017}
{Metzger}, B.~D. 2017, Living Reviews in Relativity, 20, 3,
  \dodoi{10.1007/s41114-017-0006-z}

\bibitem[{{Metzger}(2019)}]{Metzger2019}
---. 2019, Living Reviews in Relativity, 23, 1,
  \dodoi{10.1007/s41114-019-0024-0}

\bibitem[{{Metzger} \& {Berger}(2012)}]{Metzger2012}
{Metzger}, B.~D., \& {Berger}, E. 2012, \apj, 746, 48,
  \dodoi{10.1088/0004-637X/746/1/48}

\bibitem[{{Metzger} {et~al.}(2010){Metzger}, {Mart{\'\i}nez-Pinedo}, {Darbha},
  {Quataert}, {Arcones}, {Kasen}, {Thomas}, {Nugent}, {Panov}, \&
  {Zinner}}]{Metzger2010}
{Metzger}, B.~D., {Mart{\'\i}nez-Pinedo}, G., {Darbha}, S., {et~al.} 2010,
  \mnras, 406, 2650, \dodoi{10.1111/j.1365-2966.2010.16864.x}

\bibitem[{{Miller} {et~al.}(2020){Miller}, {Sprouse}, {Fryer}, {Ryan},
  {Dolence}, {Mumpower}, \& {Surman}}]{Miller2020}
{Miller}, J.~M., {Sprouse}, T.~M., {Fryer}, C.~L., {et~al.} 2020, \apj, 902,
  66, \dodoi{10.3847/1538-4357/abb4e3}

\bibitem[{{Miller} {et~al.}(2019){Miller}, {Ryan}, {Dolence}, {Burrows},
  {Fontes}, {Fryer}, {Korobkin}, {Lippuner}, {Mumpower}, \&
  {Wollaeger}}]{Miller2019}
{Miller}, J.~M., {Ryan}, B.~R., {Dolence}, J.~C., {et~al.} 2019, \prd, 100,
  023008, \dodoi{10.1103/PhysRevD.100.023008}

\bibitem[{{Misch} {et~al.}(2021){Misch}, {Ghorui}, {Banerjee}, {Sun}, \&
  {Mumpower}}]{Misch2021}
{Misch}, G.~W., {Ghorui}, S.~K., {Banerjee}, P., {Sun}, Y., \& {Mumpower},
  M.~R. 2021, \apjs, 252, 2, \dodoi{10.3847/1538-4365/abc41d}

\bibitem[{{Misch} {et~al.}(2020){Misch}, {Sprouse}, \& {Mumpower}}]{Misch2020}
{Misch}, G.~W., {Sprouse}, T.~M., \& {Mumpower}, M.~R. 2020, arXiv e-prints,
  arXiv:2011.11889.
\newblock \doarXiv{2011.11889}

\bibitem[{Moler \& Van~Loan(2003)}]{moler2003}
Moler, C., \& Van~Loan, C. 2003, SIAM review, 45, 3

\bibitem[{{M{\"o}ller} {et~al.}(2019){M{\"o}ller}, {Mumpower}, {Kawano}, \&
  {Myers}}]{Moller2019}
{M{\"o}ller}, P., {Mumpower}, M.~R., {Kawano}, T., \& {Myers}, W.~D. 2019,
  Atomic Data and Nuclear Data Tables, 125, 1,
  \dodoi{10.1016/j.adt.2018.03.003}

\bibitem[{{M{\"o}ller} {et~al.}(2016){M{\"o}ller}, {Sierk}, {Ichikawa}, \&
  {Sagawa}}]{Moller2016}
{M{\"o}ller}, P., {Sierk}, A.~J., {Ichikawa}, T., \& {Sagawa}, H. 2016, Atomic
  Data and Nuclear Data Tables, 109, 1, \dodoi{10.1016/j.adt.2015.10.002}

\bibitem[{{Mumpower} {et~al.}(2014){Mumpower}, {Cass}, {Passucci}, {Surman}, \&
  {Aprahamian}}]{Mumpower2014}
{Mumpower}, M., {Cass}, J., {Passucci}, G., {Surman}, R., \& {Aprahamian}, A.
  2014, AIP Advances, 4, 041009, \dodoi{10.1063/1.4867192}

\bibitem[{{Mumpower} {et~al.}(2020){Mumpower}, {Jaffke}, {Verriere}, \&
  {Randrup}}]{Mumpower2020}
{Mumpower}, M.~R., {Jaffke}, P., {Verriere}, M., \& {Randrup}, J. 2020, \prc,
  101, 054607, \dodoi{10.1103/PhysRevC.101.054607}

\bibitem[{{Mumpower} {et~al.}(2016{\natexlab{a}}){Mumpower}, {Kawano}, \&
  {M{\"o}ller}}]{Mumpower2016}
{Mumpower}, M.~R., {Kawano}, T., \& {M{\"o}ller}, P. 2016{\natexlab{a}}, \prc,
  94, 064317, \dodoi{10.1103/PhysRevC.94.064317}

\bibitem[{{Mumpower} {et~al.}(2018){Mumpower}, {Kawano}, {Sprouse}, {Vassh},
  {Holmbeck}, {Surman}, \& {M{\"o}ller}}]{Mumpower2018}
{Mumpower}, M.~R., {Kawano}, T., {Sprouse}, T.~M., {et~al.} 2018, \apj, 869,
  14, \dodoi{10.3847/1538-4357/aaeaca}

\bibitem[{{Mumpower} {et~al.}(2016{\natexlab{b}}){Mumpower}, {Surman},
  {McLaughlin}, \& {Aprahamian}}]{Mumpower2016r}
{Mumpower}, M.~R., {Surman}, R., {McLaughlin}, G.~C., \& {Aprahamian}, A.
  2016{\natexlab{b}}, Progress in Particle and Nuclear Physics, 86, 86,
  \dodoi{10.1016/j.ppnp.2015.09.001}

\bibitem[{{Neufcourt} {et~al.}(2020){Neufcourt}, {Cao}, {Giuliani},
  {Nazarewicz}, {Olsen}, \& {Tarasov}}]{Neufcourt2020}
{Neufcourt}, L., {Cao}, Y., {Giuliani}, S.~A., {et~al.} 2020, \prc, 101,
  044307, \dodoi{10.1103/PhysRevC.101.044307}

\bibitem[{Ney {et~al.}(2020)Ney, Engel, Li, \& Schunck}]{Ney2020}
Ney, E.~M., Engel, J., Li, T., \& Schunck, N. 2020, Phys. Rev. C, 102, 034326,
  \dodoi{10.1103/PhysRevC.102.034326}

\bibitem[{{Orford} {et~al.}(2018){Orford}, {Vassh}, {Clark}, {McLaughlin},
  {Mumpower}, {Savard}, {Surman}, {Aprahamian}, {Buchinger}, {Burkey},
  {Gorelov}, {Hirsh}, {Klimes}, {Morgan}, {Nystrom}, \& {Sharma}}]{Orford2018}
{Orford}, R., {Vassh}, N., {Clark}, J.~A., {et~al.} 2018, \prl, 120, 262702,
  \dodoi{10.1103/PhysRevLett.120.262702}

\bibitem[{Pusa \& Lepp{\"a}nen(2010)}]{pusa2010}
Pusa, M., \& Lepp{\"a}nen, J. 2010, Nuclear science and engineering, 164, 140

\bibitem[{{Radice} {et~al.}(2018){Radice}, {Perego}, {Hotokezaka}, {Fromm},
  {Bernuzzi}, \& {Roberts}}]{Radice2018}
{Radice}, D., {Perego}, A., {Hotokezaka}, K., {et~al.} 2018, \apj, 869, 130,
  \dodoi{10.3847/1538-4357/aaf054}

\bibitem[{{Schunck} {et~al.}(2015){Schunck}, {McDonnell}, {Higdon}, {Sarich},
  \& {Wild}}]{Schunck2015}
{Schunck}, N., {McDonnell}, J.~D., {Higdon}, D., {Sarich}, J., \& {Wild}, S.~M.
  2015, European Physical Journal A, 51, 169,
  \dodoi{10.1140/epja/i2015-15169-9}

\bibitem[{{Seeger} {et~al.}(1965){Seeger}, {Fowler}, \& {Clayton}}]{Seeger1965}
{Seeger}, P.~A., {Fowler}, W.~A., \& {Clayton}, D.~D. 1965, \apjs, 11, 121,
  \dodoi{10.1086/190111}

\bibitem[{{Shafer} {et~al.}(2016){Shafer}, {Engel}, {Fr{\"o}hlich},
  {McLaughlin}, {Mumpower}, \& {Surman}}]{Shafer2016}
{Shafer}, T., {Engel}, J., {Fr{\"o}hlich}, C., {et~al.} 2016, \prc, 94, 055802,
  \dodoi{10.1103/PhysRevC.94.055802}

\bibitem[{Sneden {et~al.}(2008)Sneden, Cowan, \& Gallino}]{Sneden2008}
Sneden, C., Cowan, J.~J., \& Gallino, R. 2008, Annual Review of Astronomy and
  Astrophysics, 46, 241, \dodoi{10.1146/annurev.astro.46.060407.145207}

\bibitem[{{Sprouse} {et~al.}(2020{\natexlab{a}}){Sprouse}, {Mumpower}, \&
  {Surman}}]{Sprouse2020b}
{Sprouse}, T.~M., {Mumpower}, M.~R., \& {Surman}, R. 2020{\natexlab{a}}, arXiv
  e-prints, arXiv:2008.06075.
\newblock \doarXiv{2008.06075}

\bibitem[{{Sprouse} {et~al.}(2020{\natexlab{b}}){Sprouse}, {Navarro Perez},
  {Surman}, {Mumpower}, {McLaughlin}, \& {Schunck}}]{Sprouse2020a}
{Sprouse}, T.~M., {Navarro Perez}, R., {Surman}, R., {et~al.}
  2020{\natexlab{b}}, \prc, 101, 055803, \dodoi{10.1103/PhysRevC.101.055803}

\bibitem[{{Spyrou} {et~al.}(2016){Spyrou}, {Liddick}, {Naqvi}, {Crider},
  {Dombos}, {Bleuel}, {Brown}, {Couture}, {Crespo Campo}, {Guttormsen},
  {Larsen}, {Lewis}, {M{\"o}ller}, {Mosby}, {Mumpower}, {Perdikakis}, {Prokop},
  {Renstr{\o}m}, {Siem}, {Quinn}, \& {Valenta}}]{Spyrou2016}
{Spyrou}, A., {Liddick}, S.~N., {Naqvi}, F., {et~al.} 2016, \prl, 117, 142701,
  \dodoi{10.1103/PhysRevLett.117.142701}

\bibitem[{{Spyrou} {et~al.}(2017){Spyrou}, {Larsen}, {Liddick}, {Naqvi},
  {Crider}, {Dombos}, {Guttormsen}, {Bleuel}, {Couture}, {Crespo Campo},
  {Lewis}, {Mosby}, {Mumpower}, {Perdikakis}, {Prokop}, {Quinn}, {Renstr{\o}m},
  {Siem}, \& {Surman}}]{Spyrou2017}
{Spyrou}, A., {Larsen}, A.~C., {Liddick}, S.~N., {et~al.} 2017, Journal of
  Physics G Nuclear Physics, 44, 044002, \dodoi{10.1088/1361-6471/aa5ae7}

\bibitem[{Sun {et~al.}(2008)Sun, Knöbel, Litvinov, Geissel, Meng, Beckert,
  Bosch, Boutin, Brandau, Chen, Cullen, Dimopoulou, Fabian, Hausmann,
  Kozhuharov, Litvinov, Mazzocco, Montes, Münzenberg, Musumarra, Nakajima,
  Nociforo, Nolden, Ohtsubo, Ozawa, Patyk, Plaß, Scheidenberger, Steck,
  Suzuki, Walker, Weick, Winckler, Winkler, \& Yamaguchi}]{Sun2008}
Sun, B., Knöbel, R., Litvinov, Y., {et~al.} 2008, Nuclear Physics A, 812, 1 ,
  \dodoi{https://doi.org/10.1016/j.nuclphysa.2008.08.013}

\bibitem[{{Surman} {et~al.}(2015){Surman}, {Mumpower}, \&
  {Aprahamian}}]{Surman2015}
{Surman}, R., {Mumpower}, M., \& {Aprahamian}, A. 2015, in Proceedings of the
  Conference on Advances in Radioactive Isotope Science (ARIS2014, 010010,
  \dodoi{10.7566/JPSCP.6.010010}

\bibitem[{{Surman} {et~al.}(2014){Surman}, {Mumpower}, {Sinclair}, {Jones},
  {Hix}, \& {McLaughlin}}]{Surman2014}
{Surman}, R., {Mumpower}, M., {Sinclair}, R., {et~al.} 2014, AIP Advances, 4,
  041008, \dodoi{10.1063/1.4867191}

\bibitem[{{Suzuki} {et~al.}(2012){Suzuki}, {Yoshida}, {Kajino}, \&
  {Otsuka}}]{Suzuki2012}
{Suzuki}, T., {Yoshida}, T., {Kajino}, T., \& {Otsuka}, T. 2012, \prc, 85,
  015802, \dodoi{10.1103/PhysRevC.85.015802}

\bibitem[{{Tang} {et~al.}(2020){Tang}, {Kay}, {Hoffman}, {Schiffer}, {Sharp},
  {Gaffney}, {Freeman}, {Mumpower}, {Arokiaraj}, {Baader}, {Butler}, {Catford},
  {de Angelis}, {Flavigny}, {Gott}, {Gregor}, {Konki}, {Labiche}, {Lazarus},
  {MacGregor}, {Martel}, {Page}, {Podoly{\'a}k}, {Poleshchuk}, {Raabe},
  {Recchia}, {Smith}, {Szwec}, \& {Yang}}]{Tang2020}
{Tang}, T.~L., {Kay}, B.~P., {Hoffman}, C.~R., {et~al.} 2020, \prl, 124,
  062502, \dodoi{10.1103/PhysRevLett.124.062502}

\bibitem[{{Tanvir} {et~al.}(2017){Tanvir}, {Levan},
  {Gonz{\'a}lez-Fern{\'a}ndez}, {Korobkin}, {Mandel}, {Rosswog}, {Hjorth},
  {D'Avanzo}, {Fruchter}, {Fryer}, {Kangas}, {Milvang-Jensen}, {Rosetti},
  {Steeghs}, {Wollaeger}, {Cano}, {Copperwheat}, {Covino}, {D'Elia}, {de Ugarte
  Postigo}, {Evans}, {Even}, {Fairhurst}, {Figuera Jaimes}, {Fontes}, {Fujii},
  {Fynbo}, {Gompertz}, {Greiner}, {Hodosan}, {Irwin}, {Jakobsson},
  {J{\o}rgensen}, {Kann}, {Lyman}, {Malesani}, {McMahon}, {Melandri},
  {O'Brien}, {Osborne}, {Palazzi}, {Perley}, {Pian}, {Piranomonte}, {Rabus},
  {Rol}, {Rowlinson}, {Schulze}, {Sutton}, {Th{\"o}ne}, {Ulaczyk}, {Watson},
  {Wiersema}, \& {Wijers}}]{Tanvir2017}
{Tanvir}, N.~R., {Levan}, A.~J., {Gonz{\'a}lez-Fern{\'a}ndez}, C., {et~al.}
  2017, \apjl, 848, L27, \dodoi{10.3847/2041-8213/aa90b6}

\bibitem[{Thomas \& Barber(1994)}]{thomas1994}
Thomas, G., \& Barber, D. 1994, Annals of Nuclear Energy, 21, 309

\bibitem[{{Timmes} {et~al.}(2019){Timmes}, {Fryer}, {Timmes}, {Hungerford},
  {Couture}, {Adams}, {Aoki}, {Arcones}, {Arnett}, {Auchettl}, {Avila},
  {Badenes}, {Baron}, {Bauswein}, {Beacom}, {Blackmon}, {Blondin}, {Bloser},
  {Boggs}, {Boss}, {Brandt}, {Bravo}, {Brown}, {Brown}, {Bruenn},
  {Budtz-J{\o}rgensen}, {Burns}, {Calder}, {Caputo}, {Champagne}, {Chevalier},
  {Chieffi}, {Chipps}, {Cinabro}, {Clarkson}, {Clayton}, {Coc}, {Connolly},
  {Conroy}, {C{\^o}t{\'e}}, {Couch}, {Dauphas}, {deBoer}, {Deibel},
  {Denisenkov}, {Desch}, {Dessart}, {Diehl}, {Doherty}, {Dom{\'\i}nguez},
  {Dong}, {Dwarkadas}, {Fan}, {Fields}, {Fields}, {Filippenko}, {Fisher},
  {Foucart}, {Fransson}, {Fr{\"o}hlich}, {Fuller}, {Gibson}, {Giryanskaya},
  {G{\"o}rres}, {Goriely}, {Grebenev}, {Grefenstette}, {Grohs}, {Guillochon},
  {Harpole}, {Harris}, {Harris}, {Harrison}, {Hartmann}, {Hashimoto}, {Heger},
  {Hernanz}, {Herwig}, {Hirschi}, {Hix}, {H{\"o}flich}, {Hoffman}, {Holcomb},
  {Hsiao}, {Iliadis}, {Janiuk}, {Janka}, {Jerkstrand}, {Johns}, {Jones},
  {Jos{\'e}}, {Kajino}, {Karakas}, {Karpov}, {Kasen}, {Kierans}, {Kippen},
  {Korobkin}, {Kobayashi}, {Kozma}, {Krot}, {Kumar}, {Kuvvetli}, {Laird},
  {Laming}, {Larsson}, {Lattanzio}, {Lattimer}, {Leising}, {Lennarz}, {Lentz},
  {Limongi}, {Lippuner}, {Livne}, {Lloyd-Ronning}, {Longland}, {Lopez},
  {Lugaro}, {Lutovinov}, {Madsen}, {Malone}, {Matteucci}, {McEnery}, {Meisel},
  {Messer}, {Metzger}, {Meyer}, {Meynet}, {Mezzacappa}, {Miller}, {Miller},
  {Milne}, {Misch}, {Mitchell}, {M{\"o}sta}, {Motizuki}, {M{\"u}ller},
  {Mumpower}, {Murphy}, {Nagataki}, {Nakar}, {Nomoto}, {Nugent}, {Nunes},
  {O'Shea}, {Oberlack}, {Pain}, {Parker}, {Perego}, {Pignatari}, {Pinedo},
  {Plewa}, {Poznanski}, {Priedhorsky}, {Pritychenko}, {Radice}, {Ramirez-Ruiz},
  {Rauscher}, {Reddy}, {Rehm}, {Reifarth}, {Richman}, {Ricker}, {Rijal},
  {Roberts}, {R{\"o}pke}, {Rosswog}, {Ruiter}, {Ruiz}, {Savin}, {Schatz},
  {Schneider}, {Schwab}, {Seitenzahl}, {Shen}, {Siegert}, {Sim}, {Smith},
  {Smith}, {Smith}, {Sollerman}, {Sprouse}, {Spyrou}, {Starrfield}, {Steiner},
  {Strong}, {Sukhbold}, {Suntzeff}, {Surman}, {Tanimori}, {The}, {Thielemann},
  {Tolstov}, {Tominaga}, {Tomsick}, {Townsley}, {Tsintari}, {Tsygankov},
  {Vartanyan}, {Venters}, {Vestrand}, {Vink}, {Waldman}, {Wang}, {Wang},
  {Warren}, {West}, {Wheeler}, {Wiescher}, {Winkler}, {Winter}, {Wolf},
  {Woolf}, {Woosley}, {Wu}, {Wrede}, {Yamada}, {Young}, {Zegers}, {Zingale}, \&
  {Portegies Zwart}}]{RA2020}
{Timmes}, F., {Fryer}, C., {Timmes}, F., {et~al.} 2019, \baas, 51, 2.
\newblock \doarXiv{1902.02915}

\bibitem[{Tsunoda {et~al.}(2020)Tsunoda, Otsuka, Takayanagi, Shimizu, Suzuki,
  Utsuno, Yoshida, \& Ueno}]{Tsunoda2020}
Tsunoda, N., Otsuka, T., Takayanagi, K., {et~al.} 2020, \nat, 587, 66–71,
  \dodoi{10.1038/s41586-020-2848-x}

\bibitem[{{van der Walt} {et~al.}(2011){van der Walt}, {Colbert}, \&
  {Varoquaux}}]{VanDerWalt2011}
{van der Walt}, S., {Colbert}, S.~C., \& {Varoquaux}, G. 2011, Computing in
  Science and Engineering, 13, 22, \dodoi{10.1109/MCSE.2011.37}

\bibitem[{{Vassh} {et~al.}(2020){Vassh}, {Mumpower}, {McLaughlin}, {Sprouse},
  \& {Surman}}]{Vassh2020}
{Vassh}, N., {Mumpower}, M.~R., {McLaughlin}, G.~C., {Sprouse}, T.~M., \&
  {Surman}, R. 2020, \apj, 896, 28, \dodoi{10.3847/1538-4357/ab91a9}

\bibitem[{{Vassh} {et~al.}(2019){Vassh}, {Vogt}, {Surman}, {Randrup},
  {Sprouse}, {Mumpower}, {Jaffke}, {Shaw}, {Holmbeck}, {Zhu}, \&
  {McLaughlin}}]{Vassh2019}
{Vassh}, N., {Vogt}, R., {Surman}, R., {et~al.} 2019, Journal of Physics G
  Nuclear Physics, 46, 065202, \dodoi{10.1088/1361-6471/ab0bea}

\bibitem[{{Vilen} {et~al.}(2020){Vilen}, {Kelly}, {Kankainen}, {Brodeur},
  {Aprahamian}, {Canete}, {de Groote}, {de Roubin}, {Eronen}, {Jokinen},
  {Moore}, {Mumpower}, {Nesterenko}, {O'Brien}, {Perdomo}, {Penttil{\"a}},
  {Reponen}, {Rinta-Antila}, \& {Surman}}]{Vilen2020}
{Vilen}, M., {Kelly}, J.~M., {Kankainen}, A., {et~al.} 2020, \prc, 101, 034312,
  \dodoi{10.1103/PhysRevC.101.034312}

\bibitem[{Virtanen {et~al.}(2020)Virtanen, Gommers, Oliphant, Haberland, Reddy,
  Cournapeau, Burovski, Peterson, Weckesser, Bright, {van der Walt}, Brett,
  Wilson, Millman, Mayorov, Nelson, Jones, Kern, Larson, Carey, Polat, Feng,
  Moore, {VanderPlas}, Laxalde, Perktold, Cimrman, Henriksen, Quintero, Harris,
  Archibald, Ribeiro, Pedregosa, {van Mulbregt}, \& {SciPy 1.0
  Contributors}}]{Virtanen2020}
Virtanen, P., Gommers, R., Oliphant, T.~E., {et~al.} 2020, Nature Methods, 17,
  261, \dodoi{10.1038/s41592-019-0686-2}

\bibitem[{Wanajo {et~al.}(2014)Wanajo, Sekiguchi, Nishimura, Kiuchi, Kyutoku,
  \& Shibata}]{Wanajo2014}
Wanajo, S., Sekiguchi, Y., Nishimura, N., {et~al.} 2014, The Astrophysical
  Journal, 789, L39, \dodoi{10.1088/2041-8205/789/2/l39}

\bibitem[{Wang {et~al.}(2017)Wang, Audi, Kondev, Huang, Naimi, \&
  Xu}]{Wang2017}
Wang, M., Audi, G., Kondev, F.~G., {et~al.} 2017, Chinese Phys. C, 41, 030003

\bibitem[{{Wang} {et~al.}(2020){Wang}, {N3AS Collaboration}, {Vassh}, {FIRE
  Collaboration}, {Sprouse}, {Mumpower}, {Vogt}, {Randrup}, \&
  {Surman}}]{Wang2020}
{Wang}, X., {N3AS Collaboration}, {Vassh}, N., {et~al.} 2020, \apjl, 903, L3,
  \dodoi{10.3847/2041-8213/abbe18}

\bibitem[{{Ward} \& {Fowler}(1980)}]{Ward1980}
{Ward}, R.~A., \& {Fowler}, W.~A. 1980, \apj, 238, 266, \dodoi{10.1086/157983}

\bibitem[{Wilson {et~al.}(1998)Wilson, England, \& Van~Riper}]{wilson1998}
Wilson, W., England, T., \& Van~Riper, K. 1998, in Proceedings of the 4th
  Workshop on Simulating Accelerator Radiation Environments, September 13-16,
  1998, Knoxville, Tennessee, USA, 69--79

\bibitem[{{Wollaeger} {et~al.}(2018){Wollaeger}, {Korobkin}, {Fontes},
  {Rosswog}, {Even}, {Fryer}, {Sollerman}, {Hungerford}, {van Rossum}, \&
  {Wollaber}}]{Wollaeger2018}
{Wollaeger}, R.~T., {Korobkin}, O., {Fontes}, C.~J., {et~al.} 2018, \mnras,
  478, 3298, \dodoi{10.1093/mnras/sty1018}

\bibitem[{{Wu} {et~al.}(2020){Wu}, {Nishimura}, {M{\"o}ller}, {Mumpower},
  {Lozeva}, {Moon}, {Odahara}, {Baba}, {Browne}, {Daido}, {Doornenbal}, {Fang},
  {Haroon}, {Isobe}, {Jung}, {Lorusso}, {Moon}, {Patel}, {Rice}, {Sakurai},
  {Shimizu}, {Sinclair}, {S{\"o}derstr{\"o}m}, {Sumikama}, {Watanabe}, {Xu},
  {Yagi}, {Yokoyama}, {Ahn}, {Bello Garrote}, {Daugas}, {Didierjean}, {Fukuda},
  {Inabe}, {Ishigaki}, {Kameda}, {Kojouharov}, {Komatsubara}, {Kubo}, {Kurz},
  {Kwon}, {Morimoto}, {Murai}, {Nishibata}, {Schaffner}, {Sprouse}, {Suzuki},
  {Takeda}, {Tanaka}, {Tshoo}, \& {Wakabayashi}}]{Wu2020}
{Wu}, J., {Nishimura}, S., {M{\"o}ller}, P., {et~al.} 2020, \prc, 101, 042801,
  \dodoi{10.1103/PhysRevC.101.042801}

\bibitem[{Wu {et~al.}(2019)Wu, Barnes, Mart\'{\i}nez-Pinedo, \&
  Metzger}]{Wu2019}
Wu, M.-R., Barnes, J., Mart\'{\i}nez-Pinedo, G., \& Metzger, B.~D. 2019, Phys.
  Rev. Lett., 122, 062701, \dodoi{10.1103/PhysRevLett.122.062701}

\bibitem[{Yamamoto {et~al.}(2007)Yamamoto, Tatsumi, \& Sugimura}]{yamamoto2007}
Yamamoto, A., Tatsumi, M., \& Sugimura, N. 2007, Journal of nuclear science and
  technology, 44, 147

\bibitem[{{Zhi} {et~al.}(2013){Zhi}, {Caurier}, {Cuenca-Garc{\'\i}a},
  {Langanke}, {Mart{\'\i}nez-Pinedo}, \& {Sieja}}]{Zhi2013}
{Zhi}, Q., {Caurier}, E., {Cuenca-Garc{\'\i}a}, J.~J., {et~al.} 2013, \prc, 87,
  025803, \dodoi{10.1103/PhysRevC.87.025803}

\bibitem[{{Zhu} {et~al.}(2018){Zhu}, {Wollaeger}, {Vassh}, {Surman}, {Sprouse},
  {Mumpower}, {M{\"o}ller}, {McLaughlin}, {Korobkin}, {Kawano}, {Jaffke},
  {Holmbeck}, {Fryer}, {Even}, {Couture}, \& {Barnes}}]{Zhu2018}
{Zhu}, Y., {Wollaeger}, R.~T., {Vassh}, N., {et~al.} 2018, \apjl, 863, L23,
  \dodoi{10.3847/2041-8213/aad5de}

\bibitem[{{Zhu} {et~al.}(2020){Zhu}, {Lund}, {Barnes}, {Sprouse}, {Vassh},
  {McLaughlin}, {Mumpower}, \& {Surman}}]{Zhu2020}
{Zhu}, Y.~L., {Lund}, K., {Barnes}, J., {et~al.} 2020, arXiv e-prints,
  arXiv:2010.03668.
\newblock \doarXiv{2010.03668}

\end{thebibliography}
\bibliographystyle{aasjournal}

\appendix
In this appendix, we report snapshots of nuclear abundances following the radioactive decay of an $r$-process composition which reproduces the isotopic solar $r$-process abundances of \cite{Arnould2007} in the mass range ${69 \leq A \leq 204}$. The initial distribution of nuclei is constrained to lie along the one neutron dripline as predicted by the FRDM2012 mass table \citep{Moller2016}. As demonstrated in Section~\ref{sec:results}, the abundances are robust with respect to substantial perturbation of the initial composition for the times selected for this appendix, which range between $15$ minutes and $10$ years. Abundances, $Y$, are normalized such that $\sum_i \left( Y_i \cdot A_i \right) = 1,$ where the summation over $i$ runs over nuclei in the mass range ${69 \leq A \leq 204}$, and $Y_i$ and $A_i$ are the corresponding abundance and mass number, respectively. For use in applications where only a subset of the total $r$-process nuclei are produced, e.g. in consideration of a weak $r$ process, the corresponding abundance evolution may be inferred by restricting our reported abundances to the corresponding mass range and renormalizing those abundances as appropriate.       
\input{table.tex}

\

\end{document}